\documentstyle[sprocl,epsf]{article}
\def\Journal#1#2#3#4{{#1} {\bf #2}, #3 (#4)}

\def\NPB{{\em Nucl. Phys.} B}
\def\PLB{{\em Phys. Lett.}  B}
\def\PRL{\em Phys. Rev. Lett.} 
\def\PRD{{\em Phys. Rev.} D}
\def\ZPC{{\em Z. Phys.} C}
\def\Im{\rm Im}
\def\Re{\rm Re}

\def\ss{\scriptsize}

\def\be{\begin{equation}}
\def\ee{\end{equation}}
\def\bea{\begin{eqnarray}}
\def\eea{\end{eqnarray}}
\begin{document}
\begin{flushright}
TUM-HEP-290/97\\
July 1997\\
\end{flushright}
\vskip 1cm
\title{CP Violation in Low-Energy SUSY \footnote{To appear
in ``Perspectives on Supersymmetry'',
G.L. Kane editor, World Scientific, Singapore, 1997.}}
\author{ A. MASIERO }
\address{SISSA -- ISAS, Trieste and Dip. di Fisica, Universit\`a di Perugia
and INFN, Sezione di Perugia, Via Pascoli, I-06100 Perugia, Italy}
\author{ L. SILVESTRINI }
\address{Physik Department, Technische Universit\"{a}t M\"{u}nchen, D-85748
Garching, Germany}  
\maketitle\abstracts{
We discuss CP violation in the context of the minimal SUSY 
extension of the standard model as well as in a generic 
low-energy SUSY model. After analyzing the SUSY contributions to the 
observed CP violation in the kaon system, we emphasize the prospects for 
disentangling SUSY loop contributions from pure SM effects in $B$ physics.
}
\section{Introduction}
\label{sec:intro}
CP violation has major potentialities to exhibit manifestations of new physics
beyond the Standard Model (SM). 
Indeed, it is quite a general feature that new physics possesses 
new CP violating phases in addition to the 
Cabibbo-Kobayashi-Maskawa (CKM) phase $\left(\delta_{CKM}\right)$
or, even in those cases where this does not occur, $\delta_{CKM}$
shows up in interactions of the new particles, hence with potential departures
from the SM expectations. Moreover, although the SM is able to account for the
observed CP violation in the kaon system, we cannot say that we have tested so
far the SM predictions for CP violation. The detection of CP violation in $B$
physics will constitute a crucial test of the standard CKM picture within the
SM. Again, on general grounds, we expect new physics to provide departures from
the SM CKM scenario for CP violation in $B$ physics. A final remark on reasons
that make us optimistic in having new physics playing a major role in CP
violation concerns the matter-antimatter asymmetry in the Universe. Starting
from a baryon-antibaryon symmetric Universe, the SM is unable to account for
the observed baryon asymmetry. The presence of new CP-violating contributions
when one goes beyond the SM looks crucial to produce an efficient mechanism for
the generation of a satisfactory $\Delta$B asymmetry.

The above considerations apply well to the new physics represented by
low-energy supersymmetric extensions of the SM. Indeed, as we will see below,
supersymmetry (SUSY) introduces CP-violating phases in addition to
$\delta_{CKM}$ and, even if one envisages particular situations
where such extra-phases vanish, the phase $\delta_{CKM}$ itself
leads to new CP-violating contributions in processes where SUSY particles are
exchanged. CP violation in $B$ decays has all potentialities to exhibit
departures from the SM CKM picture in low-energy SUSY extensions, although, as
we will discuss, the detectability of such deviations strongly depends on the
regions of the SUSY parameter space under consideration.

The discussion will proceed in two steps. In the first part we will briefly
review the status of CP violation in the context of the minimal SUSY extension
of the SM (MSSM). As interesting as this particular SUSY extension is, in terms
of economy of parameters and, hence, predictivity, it is becoming clearer and
clearer that the MSSM is based upon drastic assumptions of simplification which
do not find so far a strong theoretical motivation. Therefore we will devote
much of our attention to the analysis of a generic low-energy SUSY extension of
the SM without any specific assumption on the large SUSY parameter space. In
this context the main point is the identification of the most promising
processes in $B$ physics for a manifestation of the SUSY CP violation
compatibly with all the stringent constraints which come from the existing data
on FCNC phenomena.

\section{CP violation in the MSSM}
\label{sec:MSSM}

In the MSSM two new ``genuine" SUSY CP-violating phases are present. They
originate from the SUSY parameters $\mu$, $M$, $A$ and $B$. The first of these
parameters is the dimensionful coefficient of the $H_u H_d$ term of the
superpotential. The remaining three parameters are present in the sector that
softly breaks the N=1 global SUSY. $M$ denotes the common value of the gaugino
masses, $A$ is the trilinear scalar coupling, while $B$ denotes the bilinear
scalar coupling. In our notation all these three parameters are
dimensionful. The simplest way to see which combinations of the phases of these
four parameters are physical~\cite{Dugan} is to notice that for vanishing
values of $\mu$,  $M$, $A$ and $B$ the theory possesses two additional
symmetries.\cite{Dimopoulos} Indeed, letting $B$ and $\mu$ vanish, a $U(1)$
Peccei-Quinn symmetry originates, which in particular rotates $H_u$ and $H_d$.
If $M$, $A$ and $B$ are set to zero, the Lagrangian acquires a continuous
$U(1)$ $R$ symmetry. Then we can consider  $\mu$,  $M$, $A$ and $B$ as spurions
which break the $U(1)_{PQ}$ and $U(1)_R$ symmetries. In this way the question
concerning the number and nature of the meaningful phases translates into the
problem of finding the independent combinations of the four parameters which
are invariant under $U(1)_{PQ}$ and $U(1)_R$ and determining their independent
phases. There are three such independent combinations, but only two of their
phases are independent. We use here the commonly adopted choice:
\begin{equation}
  \label{MSSMphases}
  \Phi_A = {\rm arg}\left( A^* M\right), \qquad 
  \Phi_B = {\rm arg}\left( B^* M\right).
\end{equation}
The main constraints on $\Phi_A$ and $\Phi_B$ come from their contribution to
the electric dipole moments of the neutron and of the electron. For instance,
the effect of $\Phi_A$ and $\Phi_B$ on the electric and chromoelectric dipole
moments of the light quarks ($u$, $d$, $s$) lead to a contribution to $d^e_N$ of
order~\cite{EDMN}  
\begin{equation}
  \label{EDMNMSSM}
  d^e_N \sim 2 \left( \frac{100 {\rm GeV}}{\tilde{m}}\right)^2 \sin \Phi_{A,B}
  \times 10^{-23} {\rm e\,\,cm},
\end{equation}
where $\tilde{m}$ here denotes a common mass for squarks and gluinos. The
present experimental bound, $d^e_N < 1.1 \times 10^{-25}$ e cm, implies that
$\Phi_{A,B}$ should be $<10^{-2}$, unless one pushes SUSY masses up to O(1
TeV). A possible caveat to such an argument calling for a fine-tuning of
$\Phi_{A,B}$ is that uncertainties in the estimate of the hadronic matrix
elements could relax the severe bound in eq.~(\ref{EDMNMSSM}).\cite{Ellis} 

In view of the previous considerations most authors dealing with the MSSM
prefer to simply put $\Phi_A$ and $\Phi_B$ equal to zero. Actually, one may
argue in favour of this choice by considering the soft breaking sector of the
MSSM as resulting from SUSY breaking mechanisms which force $\Phi_A$ and
$\Phi_B$ to vanish. For instance, it is conceivable that both $A$ and $M$
originate from one same source of $U(1)_R$ breaking. Since $\Phi_A$ ``measures"
the relative phase of $A$ and $M$, in this case it would ``naturally"vanish. In
some specific models it has been shown~\cite{Dine} that through an analogous
mechanism also $\Phi_B$ may vanish. 

If $\Phi_A=\Phi_B=0$, then the novelty of SUSY in CP violating contributions
merely arises from the presence of the CKM phase in loops where SUSY particles
run.\cite{CPSUSY} The crucial point is that the usual GIM suppression, which
plays a major role in evaluating $\varepsilon$ and $\varepsilon^\prime$ in the
SM, in the MSSM case is replaced by a super-GIM cancellation which has the same
``power" of suppression as the original GIM.\cite{chapterFCNC} Again, also in
the MSSM, as it is the case in the SM, the smallness of $\varepsilon$ and
$\varepsilon^\prime$ is guaranteed not by the smallness of $\delta_{CKM}$, but 
rather by the small CKM angles and/or small Yukawa couplings. By the same
token, we do not expect any significant departure of the MSSM from the SM
predictions also concerning CP violation in $B$ physics. As a matter of fact,
given the large lower bounds on squark and gluino masses, one expects
relatively tiny contributions of the SUSY loops in $\varepsilon$ or
$\varepsilon^\prime$ in comparison with the normal $W$ loops of the SM. Let us
be more detailed on this point.

In SUSY, FCNC CP-violating phenomena arise at one-loop~\footnote{In SUSY models
with broken $R$-parity even tree-level contributions may be present.} through
the exchange of: i)$W$ and quarks (SM contribution); ii)charginos and squarks;
iii) charged Higgs and quarks and iv) gluinos (or, more generally, neutralinos)
and squarks. Here we are considering the low-energy minimal SUSY extension of
the SM where universality of the soft-breaking terms is assumed. This implies
that at the tree level no flavour mixing in the squark mass matrices is present
when one goes to the physical quark basis. Hence, no tree-level flavour
changing $q-\tilde q- \tilde{g}$ vertices exist. However, radiative corrections
or, equivalently, the running from the large supergravity scale down to the
electroweak scale induces a mismatch in the simultaneous diagonalization of $q$
and $\tilde q$ mass matrices, hence producing flavour changes at the $q-\tilde
q- \tilde{g}$ vertices. As detailed in the chapter on FCNC in SUSY, the amount
of this mismatch in the MSSM is again proportional to a very good approximation
to the CKM angles and the Yukawa couplings. Due to the large gluino and squark
masses (their lower bound is almost 200 GeV), in spite of the presence of the
strong coupling, it has been shown~\cite{bertolini} that the gluino exchange
(or, for that matter, any neutralino exchange) is subleading with respect to
chargino $\left(\chi^\pm\right)$ and charged Higgs $\left(H^\pm\right)$
exchanges. Hence, when dealing with CP violation in the MSSM with exact
universality and $\Phi_A=\Phi_B=0$, one can confine the analysis to $\chi^\pm$
and $H^\pm$ loops. If one takes all squarks to be degenerate in mass and
heavier than $\sim 200$ GeV, then $\chi^\pm-\tilde q$ loops are obviously
severely penalized with respect to the SM $W-q$ loops (remember that at the
vertices the same CKM angles occur in both cases).

The only chance for the MSSM to produce some sizeable departure from the SM
situation in CP violation is in the particular region of the parameter space
where one has light $\tilde q$, $\chi^\pm$ and/or $H^\pm$. We have already seen
in the previous chapters that the best candidate (indeed the only one unless
$\tan \beta \sim m_t/m_b$) for a light $\tilde q$ is the stop. Hence one can
ask the following question: can the MSSM present some novelties in CP-violating
phenomena when we consider $\chi^+ - \tilde t$ loops with light $ \tilde t$, 
$\chi^+$ and/or $H^+$?

Several analyses in the literature tackle the above question or, to be more
precise, the more general problem of the effect of light $\tilde t$ and $\chi^+$
on FCNC processes.\cite{refmis}$^{\!-\,}$\cite{mpr} A first important
observation concerns the 
relative sign of the $W-t$ loop with respect to the  $\chi^+ - \tilde t$ and 
$H^+ - t$ contributions. As it is well known, the latter contribution always
interferes positively with the SM one. Interestingly enough, in the region of
the MSSM parameter space that we consider here, also the $\chi^+ - \tilde t$
contribution constructively interferes with the SM contribution. The second
point regards the composition of the lightest chargino, i.e. whether the
gaugino or higgsino component prevails. This is crucial since the light stop is
predominantly $\tilde t_R$ and, hence, if the lightest chargino is mainly a
wino then it couples to $\tilde t_R$ mostly through the $LR$ mixing in the stop
sector. Consequently, a suppression in the contribution to box diagrams going
as $\sin^4 \theta_{LR}$ is present ($\theta_{LR}$ denotes the mixing angle
between 
the lighter and heavier stops). On the other hand, if the lightest chargino is
predominantly a higgsino (i.e. $M_2 \gg \mu$ in the chargino mass matrix), then
the $\chi^+-$lighter $\tilde t$ contribution grows. In this case contributions
$\propto \theta_{LR}$ become negligible and, moreover, it can be shown that
they are independent on the sign of $\mu$. A detailed study is provided in
ref.~\cite{mpr} For instance, for $M_2/\mu=10$ they find that the
inclusion of 
the SUSY contribution to the box diagrams doubles the usual SM contribution for
values of the lighter $\tilde t$ mass up to $100-120$ GeV, using $\tan \beta
=1.8$, $M_{H^+}=100$ TeV, $m_\chi=90$ GeV and the mass of the heavier $\tilde
t$ of 250 GeV. However, if $m_\chi$ is pushed up to 300 GeV, the  $\chi^+ -
\tilde t$ loop yields a contribution which is roughly 3 times less than in the
case $m_\chi=90$ GeV, hence leading to negligible departures from the SM
expectation. In the cases where the SUSY contributions are sizeable, one
obtains relevant restrictions on the $\rho$ and $\eta$ parameters of the CKM
matrix by making a fit of the parameters $A$, $\rho$ and $\eta$ of the CKM
matrix and of the total loop contribution to the experimental values of
$\varepsilon_K$ and $\Delta M_{B_d}$. For instance, in the above-mentioned
case in which the SUSY loop contribution equals the SM $W-t$ loop, hence giving
a total loop contribution which is twice as large as in the pure SM case,
combining the $\varepsilon_K$ and $\Delta M_{B_d}$ constraints leads to a
region in the $\rho-\eta$ plane with $0.15<\rho<0.40$ and $0.18<\eta<0.32$,
excluding negative values of $\rho$.

In conclusion, the situation concerning CP violation in the MSSM case with
$\phi_A=\phi_B=0$ and exact universality in the soft-breaking sector can be
summarized in the following way: the MSSM does not lead to any significant
deviation from the SM expectation for CP-violating phenomena as $d_N^e$,
$\varepsilon$, $\varepsilon^\prime$ and CP violation in $B$ physics; the only
exception to this statement concerns a small portion of the MSSM parameter space
where a very light $\tilde t$ ($m_{\tilde t} < 100$ GeV) and $\chi^+$
($m_\chi \sim 90$ GeV) are present. In this latter particular situation
sizeable SUSY contributions to $\varepsilon_K$ are possible and, consequently,
major restrictions in the $\rho-\eta$ plane can be inferred. Obviously, CP
violation in $B$ physics becomes a crucial test for this MSSM case with very
light $\tilde t$ and $\chi^+$. Interestingly enough, such low values of SUSY
masses are at the border of the detectability region at LEP II.

\section{CP violation in general SUSY extensions of the SM}
\label{sec:gen}

We now move to the more interesting case where one envisages a generic
SUSY extension of the SM without making the quite drastic assumptions entering
the construction of the MSSM. In previous chapters we have already seen that
FCNC processes severely limit the amount of non-universality in the $\tilde q$
sector. The questions we tackle in this chapter concerning a general SUSY
extension of the SM are: i) what are the constraints imposed on $\tilde q$ mass
matrices by $\varepsilon$, $\varepsilon^\prime$ and $d_N^e$? ii) is the SUSY
contribution to CP violation in kaon physics of the superweak or milliweak
kind? iii) given the constraints from CP conserving FCNC processes involving
$B$ mesons, can we still hope that low-energy SUSY manifests itself through
sizeable departures from the SM expectations in CP violating $B$ decays?
Obviously the latter question, in particular, seems of great relevance in a
moment when new $B$-factories and $B$-dedicated programs at existing
accelerators are going to become operative very soon.

To answer the above questions we use a model-independent parameterization of
FCNC and CP-violating contributions which makes use of the so-called super-CKM
basis~\cite{mins} 
where $q-\tilde q- \tilde{g}$ couplings are diagonal in flavour and all
the flavour changing effects are due to the non-diagonality of the $\tilde q$
mass matrices. As long as the ratio of the off-diagonal entries over an average
$\tilde q$ mass remains a small parameter, the first term of the expansion which
is obtained by an off-diagonal mass insertion in the $\tilde q$ propagators
represents a suitable approximation. This method avoids the specific knowledge
of the $\tilde f$ mass matrices and allows to
account for the FCNC $\tilde g$ and $\chi^\pm$ loops. Notice that, for a given
process, the mass insertions refer to squarks of different electric charge when
$\tilde g$ and $\chi^\pm$ are exchanged in the loops and that the relative size
and sign of such different mass insertions is undetermined as long as one keeps
with generic SUSY extensions of the SM.\footnote{Except for $LL$ mass insertions
(see ref.~\cite{mpr} for details on this point).} 
Hence one cannot be conclusive about
the interplay of  $\tilde g$ and $\chi^\pm$ exchanges to a given FCNC or CP
violating process, but one should consider all the various possibilities taking
into account the full set of processes which constrain such mass
insertions. For the sake of our discussion here it is enough to concentrate on
one of the two contributions, for instance the SUSY loops with  $\tilde g$
running inside. In the approach to a generic SUSY extension we expect similar
contributions from $\tilde g$ and  $\chi^\pm$ exchanges, unless one invokes
particular regions of the SUSY parameter space where $\chi^\pm$ are very light.
Indeed, here, differently from the very constrained case of the MSSM, we
generally obtain severe constraints on the FC mass insertions even for
conspicuous values of the $\tilde g$, $\chi^\pm$ and $\tilde q$ masses. Gluino
exchange is enhanced with respect to  $\chi^\pm$ exchange by powers of $\left(
\alpha_s/\alpha_W\right)$ couplings, but, on the other hand, on quite general 
grounds
we expect  $\chi^\pm$ to be lighter than $\tilde g$ and hence a factor $\left(
m_\chi/m_{\tilde g}\right)^2$ can penalize the $\tilde g$ exchange
contribution. 
With that in mind, and barring significant cancellations between the  $\tilde
g$ and $\chi^\pm$ contributions, it makes sense to focus just on the $\tilde
g$ contribution to CP-violating processes.

We briefly recall here the main ingredients for our analysis in the framework
of the mass-insertion approximation in the super-CKM basis 
(this is useful also to
establish our notation). There exist four different $\Delta$ mass insertions 
connecting flavours $i$ and $j$ along a sfermion propagator: 
$\left(\Delta_{ij}\right)_{LL}$, 
$\left(\Delta_{ij}\right)_{RR}$, $\left(\Delta_{ij}\right)_{LR}$ and 
$\left(\Delta_{ij}\right)_{RL}$. The indices $L$ and $R$ refer to the 
helicity of 
the
fermion partners. The size of these $\Delta$'s can be quite different. For
instance, it is well known that in the MSSM case only the $LL$ mass insertion
can change flavour, while all the other three above mass insertions are flavour
conserving, i.e. they have $i=j$. In this case to realize a $LR$ or $RL$ 
flavour
change one needs a double mass 
insertion with the flavour changed solely in a $LL$
mass insertion and a subsequent flavour-conserving $LR$ mass insertion. 
Even worse
is the case of a FC $RR$ transition: in the MSSM this can be accomplished only
through a laborious set of three mass insertions, two flavour-conserving $LR$
transitions and an $LL$ FC insertion. Notice also that generally the 
$\Delta_{LR}$
quantity does not necessarily coincide with $\Delta_{RL}$. For instance, in the
MSSM and in several other cases, one flavour-conserving mass insertion is
proportional to the mass of the corresponding right-handed fermion. Hence ,
$(\Delta_{ij})_{LR}$ and $(\Delta_{ij})_{RL}$ are proportional to the mass of 
the $i$-th
and $j$-th fermion, respectively. 
Instead of the dimensional quantities $\Delta$ it is more
useful to provide bounds making use of dimensionless quantities, $\delta$, 
that are obtained dividing the mass insertions by an average sfermion mass.

\begin{figure}   
    \begin{center}
    \epsfysize=6truecm 
    \leavevmode\epsffile{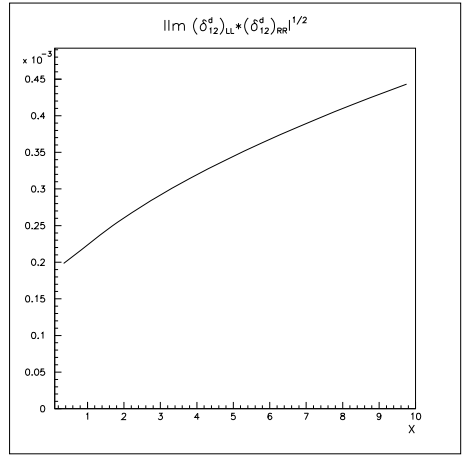}
    \end{center}
    \caption[]{The $\sqrt{\left|{\rm Im}  \left(\delta^{d}_{12} 
     \right)_{LL}^{2}\right|}$ as a function
     of $x=m_{\tilde{g}}^2/m_{\tilde{q}}^2$, for  an average squark mass 
     $m_{\tilde{q}}=500$ GeV.}
\label{fig:epllrr}
\end{figure}

We start by considering CP violation in the kaon system, i.e. $\varepsilon$ and
$\varepsilon^\prime$. In reference~\cite{GGMS} we provide a very detailed
description of how to calculate the effective hamiltonian for $\Delta S=2$ and
$\Delta S=1$ processes as well as a determination of the hadronic matrix
elements. Asking for each $\tilde g$ exchange contribution not to exceed the
experimental value $\varepsilon = 2.268 \times 10^{-3}$ we obtain bounds on the
quantities $\left\vert {\rm Im} \left(\delta^d_{12}\right)^2_{LL}
\right\vert^{1/2}$,
$\left\vert {\rm Im} \left(\delta^d_{12}\right)^2_{LR}\right\vert^{1/2}$ 
and $\left\vert {\rm
Im} \left(\delta^d_{12}\right)_{LL}\left(\delta^d_{12}\right)_{RR}
\right\vert^{1/2}$
as a function of the average squark mass $m_{\tilde q}$, and of the ratio
$x=\left(m_{\tilde g}/m_{\tilde q}\right)^2$. In table \ref{tab:imds2} we report
such limits taking $x=0.3$, 1 and 4, for $m_{\tilde q}=500$ GeV. For different
values of $m_{\tilde q}$, the limits can be obtained multiplying the values in
table~\ref{tab:imds2} by $m_{\tilde q}$ (GeV)/500.
\begin{table}
 \begin{center}
 \caption[]{Limits on 
 $\mbox{Im}\left(\delta_{12}^{d}\right)_{AB}\left(\delta_{12}^{d}\right)_{CD}$,
 with $A,B,C,D=(L,R)$, for 
 an average squark mass $m_{\tilde{q}}=500\mbox{GeV}$ and for different values
 of  
 $x=m_{\tilde{g}}^2/m_{\tilde{q}}^2$. For different values of $m_{\tilde{q}}$, 
 the limits can be obtained multiplying the ones in the table by 
 $m_{\tilde{q}}(\mbox{GeV})/500$.}
 \label{tab:imds2}
 \begin{tabular}{|c|c|c|c|}  \hline
  $x$ &
 ${\scriptstyle\sqrt{\left|\Im  \left(\delta^{d}_{12} \right)_{LL}^{2}
\right|} }$ &
 ${\scriptstyle\sqrt{\left|\Im  \left(\delta^{d}_{12} \right)_{LR}^{2}
\right|} }$ &
 ${\scriptstyle\sqrt{\left|\Im  \left(\delta^{d}_{12} \right)_{LL}
 \left(\delta^{d}_{12}
 \right)_{RR}\right|} }$ \\
 \hline
 $
   0.3
 $ &
 $
1.5\times 10^{-3}
 $ & $
6.3\times 10^{-4}
 $ & $
2.0\times 10^{-4}
 $ \\
 $
   1.0
 $ &
 $
3.2\times 10^{-3}
 $ & $
3.5\times 10^{-4}
 $ & $
2.2\times 10^{-4}
 $ \\
 $
   4.0
 $ &
 $
7.5\times 10^{-3}
 $ & $
4.2\times 10^{-4}
 $ & $
3.2\times 10^{-4}
 $ \\ \hline
 \end{tabular}
 \end{center}
 \end{table}
In fig.~\ref{fig:epllrr} we plot the bound on $\left\vert{\rm Im}
\left(\delta^{d}_{12}  
\right)_{LL}^{2}\right\vert^{1/2}$ as a function 
of $x$ for $m_{\tilde{q}}=500{\rm GeV}$. It should be noticed that the bounds
derived from $\varepsilon$ on the imaginary parts of products of $\delta$'s are
one order of magnitude more stringent than the corresponding limits on the real
parts which are obtained from $\Delta M_K$.

\begin{figure}   
    \begin{center}
    \epsfysize=6truecm
    \leavevmode\epsffile{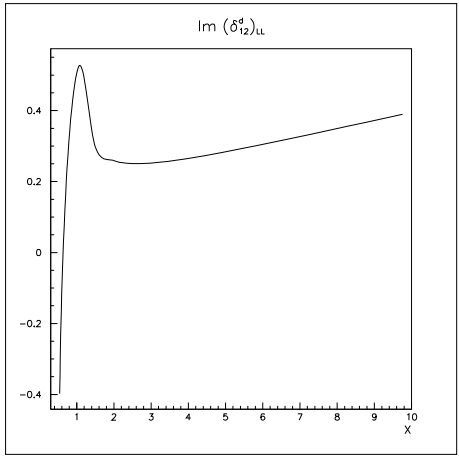}
    \end{center}
    \caption[]{The ${\rm Im}\left(\delta^d_{12}\right)_{LL}$ as a function
     of $x=m_{\tilde{g}}^2/m_{\tilde{q}}^2$, for  an average squark mass 
     $m_{\tilde{q}}=500$ GeV.}
     \label{fig:eppll}
\end{figure}

\begin{figure}   
    \begin{center}
    \epsfysize=6truecm
    \leavevmode\epsffile{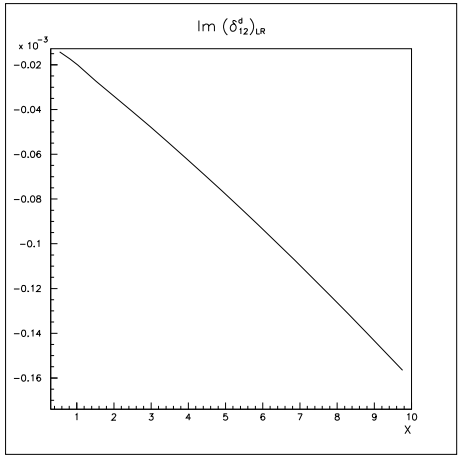}
    \end{center}
    \caption[]{The  ${\rm Im}\left(\delta^d_{12}\right)_{LR}$ as a function
     of $x=m_{\tilde{g}}^2/m_{\tilde{q}}^2$, for  an average squark mass 
     $m_{\tilde{q}}=500$ GeV.}
     \label{fig:epplr}
\end{figure}

Coming to $\Delta S=1$ processes, both superpenguin and superboxes contribute
to $\varepsilon^\prime$. It was only very 
recently~\cite{GGMS}$^{\!-\,}$\cite{GMS} that it was
realized that superboxes are at least as important as superpenguin diagrams in
contributions which proceed through a $\left(\delta^{d}_{12}\right)_{LL}$
insertion. In fig.~\ref{fig:eppll} we report the bound on ${\rm
Im}\left(\delta^d_{12}\right)_{LL}$ as a function of $x$ for $m_{\tilde q}=500$
GeV which comes from the conservative demand that
$\varepsilon^\prime/\varepsilon < 2.7 \times 10^{-3}$. The contribution of box
and penguin diagrams to the $LL$  terms have opposite signs and a sizeable
cancellation occurs for $x$ close to one, where the two contribution are of
comparable size (this explains the peak around $x=1$ in the plot of
fig.~\ref{fig:eppll}). A much more stringent limit is obtained for
$\left(\delta^{d}_{12}\right)_{LR}$ (fig.~\ref{fig:epplr}). For the $LR$
contribution only superpenguins play a relevant role. Speaking of
superpenguins, it is interesting to notice that, differently from the SM case,
the SUSY contributions are negligibly affected by electroweak penguins,
i.e. gluino-mediated $Z^0$- or $\gamma$-penguins are strongly suppressed with
respect to gluino-mediated gluon penguins.\cite{GGMS} 


In table \ref{tab:imds1} we summarize the bounds on ${\rm Im}
\left(\delta^{d}_{12}  \right)_{LL}$ and ${\rm Im}
\left(\delta^{d}_{12}  \right)_{LR}$ coming from
$\varepsilon^{\prime}/\varepsilon < 2.7 \times 10^{-3}$ for the same values of
SUSY masses chosen in table \ref{tab:imds2}. The comparison of the two tables
leads to the following two conclusions:
\begin{enumerate}
\item if we consider a SUSY extension of the SM where the $LR$ insertions are
much smaller than the $LL$ ones (this is what occurs in the MSSM, for instance),
then fulfilling the bound coming from $\varepsilon$ implies that ${\rm Im}
\left(\delta^{d}_{12}  \right)_{LL}$ is too small to provide a sizeable
contribution to $\varepsilon^{\prime}$ unless $\left(\delta^{d}_{12}
\right)_{LL}$ is almost purely imaginary (remember that $\varepsilon$ bounds
${\rm Im} \left(\delta^{d}_{12}  \right)_{LL}^2$). Hence, in this case the SUSY
contribution would be of superweak nature;
\item if, on the contrary, we have a SUSY model with sizeable $LR$ $\Delta S=1$
mass insertions, then it is possible to respect the bound from $\varepsilon$,
while obtaining a large contribution to $\varepsilon^{\prime}/\varepsilon$. In
this case we would have a SUSY milliweak contribution to CP violation. For this
to occur, we need a SUSY model where $ \left(\delta^{d}_{12}  \right)_{LR}$ is
no longer proportional to $m_s$, but rather to some much larger mass.
\end{enumerate}

 \begin{table}
 \begin{center}
  \caption[]{Limits from $\varepsilon^{\prime}/\varepsilon < 2.7 \times
  10^{-3}$  
 on  $\Im\left(\delta_{12}^{d}\right)$, for 
 an average squark mass $m_{\tilde{q}}=500\mbox{GeV}$ and for different values
 of  
 $x=m_{\tilde{g}}^2/m_{\tilde{q}}^2$. For different values of $m_{\tilde{q}}$, 
 the limits can be obtained multiplying the ones in the table by 
 $\left(m_{\tilde{q}}(\mbox{GeV})/500\right)^2$.}
 \label{tab:imds1}
\begin{tabular}{|c|c|c|}   \hline
  & & \\
  $x$ & ${\scriptstyle\left\vert\Im \left(\delta^{d}_{12}  \right)_{LL}
\right\vert} $ &
 ${\scriptstyle\left\vert\Im \left(\delta^{d}_{12}\right)  _{LR}\right\vert }$ 
\\
  & & \\\hline
 $
   0.3
 $ &
 $
1.0\times 10^{-1}
 $ & $
1.1\times 10^{-5}
 $ \\
 $
   1.0
 $ &
 $
4.8\times 10^{-1}
 $ & $
2.0\times 10^{-5}
 $ \\
 $
   4.0
 $ &
 $
2.6\times 10^{-1}
 $ & $
6.3\times 10^{-5}
 $ \\ \hline 
 \end{tabular}
 \end{center}
 \end{table}

The above latter remark would lead us to the natural conclusion that to have
sizeable SUSY contributions to $\varepsilon^{\prime}/\varepsilon$ one needs a
SUSY extension of the SM where $\tilde{q}_L-\tilde{q}_R$ transitions are no
longer proportional to $m_q$ (we remind the reader that in the MSSM a mass term
$\tilde{q}_L\tilde{q}_R^*$ receives two contributions, one proportional to the
parameter A and the other to $\mu$, but both of them are proportional to
$m_q$). However, if this enhancement occurs also for flavour-conserving
$\tilde{q}_L-\tilde{q}_R$ transitions, one may envisage some problem with the
very stringent bound on the $d^e_N$. Indeed, imposing this latter constraint
yields the following limits on ${\rm Im}\left(\delta_{11}^{d}\right)_{LR}$ for
$m_{\tilde{q}}=500$ GeV:
\begin{eqnarray}
  x &\qquad&{\rm Im}\left(\delta_{11}^{d}\right)_{LR}\nonumber \\
0.3 & \qquad & 2.4 \times 10^{-6}\nonumber \\
1.0 &  \qquad & 3.0 \times 10^{-6}\nonumber \\
4.0 &  \qquad & 5.6 \times 10^{-6}
  \label{imd11lr}
\end{eqnarray}
A quick comparison of the above numbers with the bounds on ${\rm
Im}\left(\delta_{12}^{d}\right)_{LR}$ from  $\varepsilon^{\prime}/\varepsilon$
reveals that to get a sizeable SUSY contribution to $\varepsilon^{\prime}$ we
need values of  ${\rm
Im}\left(\delta_{12}^{d}\right)_{LR}$ which exceed the bound on ${\rm
Im}\left(\delta_{11}^{d}\right)_{LR}$ arising from $d^e_N$. Obviously, strictly
speaking it is not forbidden for  ${\rm
Im}\left(\delta_{12}^{d}\right)_{LR}$ to be $\ge{\rm
Im}\left(\delta_{11}^{d}\right)_{LR}$, but certainly such a probability does
not look straightforward. In conclusion, although technically it is conceivable
that some SUSY extension may provide a large
$\varepsilon^{\prime}/\varepsilon$, it is rather difficult to imagine how to
reconcile such a large enhancement of  ${\rm
Im}\left(\delta_{12}^{d}\right)_{LR}$ with the very strong constraint on the
flavour-conserving  ${\rm
Im}\left(\delta_{11}^{d}\right)_{LR}$  from $d^e_N$.

\section{CP violation in $B$ physics}

We now move to the next frontier for testing the unitarity triangle in 
general and in particular CP violation in the SM and its SUSY extensions: $B$ 
physics. We have seen above that the transitions between 1st and 2nd 
generation 
in the down sector put severe constraints on $\Re \delta^d_{12}$ and 
$\Im \delta^d_{12}$  
quantities. To be sure, the bounds derived from $\varepsilon$ and 
$\varepsilon^\prime$ are stronger than the corresponding bounds from $\Delta 
M_K$. If the same pattern repeats itself in the transition between 3rd and 1st 
or 3rd and 2nd generation in the down sector we may expect that the constraints
inferred from $B_d - \bar{B}_d$ oscillations or $b \to s \gamma$ do not prevent
conspicuous new contributions 
 also in 
CP-violating processes in $B$ physics. We are going to see below that this is 
indeed the case and we will argue that measurements of CP asymmetries in 
several $B$-decay channels may allow to disentangle SM and SUSY contributions 
to the CP decay phase.

First, we consider the constraints on $\delta^d_{13}$ and $\delta^d_{23}$ from 
$B_d - \bar{B}_d$ and $b \to s \gamma$, respectively. From the former process 
we obtain the bounds on $\Re \left(\delta^d_{13}\right)^2_{LL}$, $\Re
\left(\delta^d_{13}\right)^2_{LR}$ and  
$\Re \left(\delta^d_{13}\right)_{LL}\left(\delta^d_{13}\right)_{RR}$ which are
reported in table~\ref{tab:bbar} for the usual three typical values of $x$ ($x
= 0.3$, 1 and 4) and for $m_{\tilde q}=500 $ GeV (the results scale with
$m_{\tilde q}({\rm GeV})/ 500 $ for different values of $m_{\tilde q}$).

\begin{table}
 \begin{center}
 \caption[]{Limits on $\mbox{Re}\left(\delta_{13}\right)_{AB}\left(
\delta_{13}\right)_{CD}$, with $A,B,C,D=(L,R)$, for an average squark mass
 $m_{\tilde{q}}=500\mbox{GeV}$ and for different values of 
 $x=m_{\tilde{g}}^2/m_{\tilde{q}}^2$. For different values of $m_{\tilde{q}}$, 
 the limits can be obtained multiplying the ones in the table by 
 $m_{\tilde{q}}(\mbox{GeV})/500$.}
 \label{tab:bbar}
 \begin{tabular}{|c|c|c|c|}  \hline 
 $x$ & $\sqrt{\left|\Re  \left(\delta^{d}_{13} \right)_{LL}^{2}\right|} $ 
 &
 $\sqrt{\left|\Re  \left(\delta^{d}_{13} \right)_{LR}^{2}\right|} $ &
 $\sqrt{\left|\Re  \left(\delta^{d}_{13} \right)_{LL}\left(\delta^{d}_{13}
 \right)_{RR}\right|} $ \\
 \hline
 $
   0.3
 $ &
 $
4.6\times 10^{-2}
 $ & $
5.6\times 10^{-2}
 $ & $
1.6\times 10^{-2}
 $ \\
 $
   1.0
 $ &
 $ 
9.8\times 10^{-2}
 $ & $
3.3\times 10^{-2}
 $ & $
1.8\times 10^{-2}
 $ \\
 $
   4.0
 $ &
 $
2.3\times 10^{-1}
 $ & $
3.6\times 10^{-2}
 $ & $
2.5\times 10^{-2}
 $ \\ \hline
 \end{tabular}
 \end{center}
 \end{table}

The radiative decay $b \to s \gamma$ constraints only the $\left\vert
\left(\delta^d_{23}\right)_{LR}\right\vert$ quantity in a significant way. For
$m_{\tilde q}=500 $ GeV we obtain bounds on $\left\vert
\left(\delta^d_{23}\right)_{LR}\right\vert$  in the range $(1.3 \div 3) \times
10^{-2} $ for $x$ varying from 0.3 to 4, respectively (the bound scales as
$m_{\tilde q}^2$). On the other hand, $b \to s \gamma$ does not limit 
 $\left\vert\left(\delta^d_{23}\right)_{LL}\right\vert$. 
 In the following, we will take 
$\left\vert\left(\delta^d_{23}\right)_{LL}\right\vert=1$ (corresponding to
$x_s=\left(\Delta M/\Gamma\right)_{B_s} > 70 $ for $m_{\tilde q}=500 $ GeV).

New physics can modify the SM predictions on  CP asymmetries in $B$
decays~\cite{rattazzi}
by changing the phase of the $B_{d}$--$\bar{B}_{d}$ mixing
and the  phase and absolute value of  the decay amplitude. 
The general SUSY extension of the SM that we discuss here affects both these 
quantities. 

The remaining part of this chapter tackles the following question of crucial
relevance in the next few years: where and how can one possibly distinguish
SUSY contributions to CP violation in $B$ decays?~\cite{cptutti} 
As we said before we want our
answer to be as general as possible, i.e. without any commitment to particular
SUSY models. Obviously, a preliminary condition to properly answer the above
question is to estimate the amount of the uncertainties of the SM predictions
for CP asymmetries in $B$ decays.

To discuss the latter above-mentioned point, 
we choose to work in the theoretical framework of ref.~\cite{cpnoi}
We use the effective Hamiltonian
(${\cal H}_{eff}$) formalism, including
LO QCD corrections;  in the numerical analysis, we use the LO SM Wilson
coefficients evaluated at $\mu=5$ GeV, as given in ref.~\cite{zeit} 
In most  of the cases, by choosing different scales (within a resonable range) 
or by using NLO Wilson coefficients,  the
results vary by about $20-30 \%$ .  This is true
with the  exception of some particular channels where uncertainties  are 
larger. The matrix elements of the operators of ${\cal H}_{eff}$
are given in terms of the following Wick contractions 
 between hadronic states: Disconnected
Emission ($DE$), Connected Emission ($CE$), Disconnected Annihilation ($DA$),
Connected Annihilation ($CA$), Disconnected Penguin ($DP$) 
and Connected Penguin
($CP$) (either for left-left ($LL$) or for left-right ($LR$)
current-current operators). Following ref.~\cite{cfms}, where
a detailed discussion can be found, 
instead of  adopting a specific model for  estimating  the different
diagrams,  we let them vary within reasonable 
ranges.  In order to illustrate the relative 
strength and variation of the different contributions, in table~\ref{tab:ampli} 
we only show, for six different cases,
 results obtained by taking the extreme values of these ranges.
 In the first column 
only $DE=DE_{LL}=DE_{LR}$ are assumed to be different from zero. 
For simplicity, unless stated otherwise,  the same numerical 
values are used for  diagrams corresponding to the insertion 
of  $LL$ or  $LR$ operators, i.e. $DE=DE_{LL}=DE_{LR}$, 
$CE=CE_{LL}=CE_{LR}$, etc.  We then consider, 
in addition to $DE$,  the $CE$ contribution  by taking
 $CE=DE/3$. 
Annihilation diagrams  are included in the third
column, where we use  $DA=0$ and $CA=1/2 DE$.\cite{cfms} 
Inspired by kaon decays, we allow for some enhancement of 
the matrix elements  of left-right ($LR$) operators and choose
$DE_{LR}=2 DE_{LL}$ and $CE_{LR}=2 CE_{LL}$  (fourth column). 
Penguin contractions, $CP$ and $DP$, 
can be interpreted  as  long-distance penguin contributions to the matrix 
elements and play an important role:  if   we  
take $CP_{LL}=CE$ and $DP_{LL}=DE$ (fifth column), in some decays 
these terms  dominate the amplitude. 
Finally, in the sixth column,  we allow for long distance
effects which might differentiate penguin contractions with up and charm quarks
in the loop, giving rise to incomplete GIM cancellations (we assume 
$\overline{DP}= DP(c) - DP(u) =  DE/3$ and 
$\overline{CP}= CP(c) - CP(u) =CE/3$). 

\begin{table}
 \begin{center}
 \caption[]{Ratios of amplitudes for exclusive $B$ decays. For each channel, 
whenever two terms  with  different CP phases contribute in the SM, 
we  give the ratio $r$ of the two amplitudes.
For each channel,
the second and third lines, where present, contain the ratios of SUSY to SM 
contributions for SUSY masses  of 250 and 500 GeV respectively.}
 \protect\label{tab:ampli}
 \begin{tabular}{|ccccccc|}
 \hline  
 \ss{Process }&\ss{ $DE$ }&\ss{ $DE+CE$ }&\ss{
 $DE+CE$ }&\ss{ $DE+CE+$ }&\ss{
 $DE+CE+$}&\ss{ $DE+CE+$}\\ \ss{
 }&\ss{   }&\ss{   }&\ss{
 $+CA$ }&\ss{ $CA+LR$ }&\ss{
 $DP+CP$ }&\ss{ $\overline{DP}+ \overline{CP}$}\\
 \hline
&\ss{--}&\ss{--}&\ss{--}& 
\ss{--}&\ss{--}&\ss{--}\\
\ss{
$B^0_d \to J/\psi K_S$}&\ss{-0.03}&\ss{0.1}&
\ss{0.1}&\ss{0.1}&\ss{0.1}&\ss{0.1}\\
 \ss{
}&\ss{-0.008}&\ss{0.02}&\ss{0.02}&
\ss{0.04}&\ss{0.02}&\ss{0.02}\\ 
\hline
&\ss{--}&\ss{--}&\ss{--}&\ss{--}&
\ss{--}&\ss{--}\\ \ss{
$B^0_d \to \phi K_S$}&\ss{0.7}&\ss{0.7}&
\ss{0.7}&\ss{0.6}&\ss{0.4}&\ss{0.4}\\ 
\ss{
}&\ss{0.2}&\ss{0.2}&\ss{0.2}&\ss{0.1}&\ss{0.1}&\ss{0.09}\\
\hline
&\ss{0.08}&\ss{-0.06}&\ss{-0.05}&\ss{-0.02}&\ss{-0.009}&\ss{-0.01}\\ \ss{
$B^0_d \to K_S \pi^0$}&\ss{0.7}&\ss{0.7}&\ss{0.6}&\ss{0.6}&\ss{0.4}&\ss{0.4}\\ 
\ss{
}&\ss{0.2}&\ss{0.2}&\ss{0.2}&\ss{0.1}&\ss{0.1}&\ss{0.09}\\ 
\hline
\ss{$B^0_d \to D^0_{CP} \pi^0$}&\ss{0.02}&\ss{0.02}&\ss{0.02}&\ss{0.02}&
\ss{0.02}&\ss{0.02}\\
\hline
&\ss{-0.6}&\ss{0.9}&\ss{-0.7}&\ss{-2.}&\ss{6.}&\ss{4.}\\ \ss{
$B^0_d \to \pi^0 \pi^0$}&\ss{0.3}&\ss{-0.07}&\ss{0.4}&\ss{-0.4}&\ss{-0.07}&
\ss{-0.06}\\ \ss{
}&\ss{0.06}&\ss{-0.02}&\ss{0.09}&\ss{-0.1}&\ss{-0.02}&\ss{-0.02}\\ 
\hline
&\ss{-0.09}&\ss{-0.1}&\ss{-0.1}&\ss{-0.3}&\ss{-0.9}&\ss{-0.8}\\ \ss{
$B^0_d \to \pi^+\pi^-$}&\ss{0.02}&\ss{0.02}&\ss{0.03}&\ss{0.09}&\ss{0.8}&
\ss{0.4}\\ \ss{
}&\ss{0.005}&\ss{0.006}&\ss{0.008}&\ss{0.02}&\ss{0.2}&\ss{0.1}\\ 
\hline
&\ss{0.03}&\ss{0.04}&\ss{0.05}&\ss{0.1}&\ss{0.3}&\ss{0.2}\\ \ss{
$B^0_d \to D^+D^-$}&\ss{-0.007}&\ss{-0.008}&\ss{-0.01}&\ss{-0.02}&\ss{-0.02}&
\ss{-0.02}\\ \ss{
}&\ss{-0.002}&\ss{-0.002}&\ss{-0.002}&\ss{-0.005}&\ss{-0.006}&\ss{-0.005}\\ 
\hline
&\ss{0}&\ss{0}&\ss{0}&\ss{0}&\ss{0.}&\ss{0.07}\\ \ss{
$B^0_d \to K^0 \bar{K}^0$}&\ss{-0.2}&\ss{-0.2}&\ss{-0.2}&\ss{-0.2}&\ss{-0.09}&
\ss{-0.08}\\ \ss{
}&\ss{-0.06}&\ss{-0.05}&\ss{-0.05}&\ss{-0.04}&\ss{-0.02}&\ss{-0.02}\\ 
\hline
&\ss{--}&\ss{--}&\ss{-0.2}&\ss{-0.4}&\ss{--}&\ss{--}\\ \ss{
$B^0_d \to K^+K^-$}&\ss{--}&\ss{--}&\ss{0.04}&\ss{0.1}&\ss{--}&\ss{--}\\ \ss{
}&\ss{--}&\ss{--}&\ss{0.01}&\ss{0.03}&\ss{--}&\ss{--}\\
\hline
&\ss{--}&\ss{--}&\ss{--}&\ss{--}&\ss{--}&\ss{--}\\ \ss{
$B^0_d \to D^0\bar{D}^0$}&\ss{--}&\ss{--}&\ss{-0.01}&\ss{-0.03}&\ss{--}&
\ss{--}\\ \ss{
}&\ss{--}&\ss{--}&\ss{-0.003}&\ss{-0.006}&\ss{--}&\ss{--}\\ 
\hline
&\ss{-0.04}&\ss{0.1}&\ss{0.1}&\ss{0.3}&\ss{0.1}&\ss{0.1}\\ \ss{
$B^0_d \to J/\psi \pi^0$}&\ss{0.007}&\ss{-0.02}&\ss{-0.02}&\ss{-0.03}&
\ss{-0.02}&\ss{-0.02}\\ \ss{
}&\ss{0.002}&\ss{-0.005}&\ss{-0.005}&\ss{-0.008}&\ss{-0.005}&\ss{-0.005}\\ 
\hline
&\ss{--}&\ss{--}&\ss{--}&\ss{--}&\ss{--}&\ss{--}\\ \ss{
$B^0_d \to \phi\pi^0$}&\ss{-0.06}&\ss{-0.1}&\ss{-0.1}&\ss{-0.1}&\ss{-0.1}&
\ss{-0.1}\\ \ss{
}&\ss{-0.01}&\ss{-0.03}&\ss{-0.03}&\ss{-0.03}&\ss{-0.03}&\ss{-0.03}\\ \hline
 \end{tabular}
 \end{center}
 \end{table}

In addition to the ratios of the different SM contributions to the decay
amplitudes given in
table~\ref{tab:ampli}, obtained letting
the matrix elements vary in the broad range defined above, we also give, in
table~\ref{tab:BR}, the branching ratios for the channels of interest to us.
These branching ratios are obtained following the approach of ref.~\cite{bkpi}
We use QCD sum rules form factors~\cite{qcdsr} to compute the factorizable $DE$
contribution, then fit $CE$ using the available data on $b \to c$ two-body
decays; $CP$ and $DP$ are extracted from the measured $B \to K \pi$ branching
ratios, $CA$ is varied between 0 and $0.5\times DE$ 
and $DA$, $\overline{DP}$ and
$\overline{CP}$ are set to zero.  The range of values in table~\ref{tab:BR}
corresponds to the variation of the CKM angles in the presently allowed range
and to the inclusion of the contributions proportional to $CP$ and $DP$
(see ref.~\cite{bkpi} for further details).   

\begin{table}
 \begin{center}
 \caption[]{Branching ratios for $B$ decays.}
 \label{tab:BR}
 \begin{tabular}{|cc|}
 \hline 
 Channel&BR $\times 10^{5}$ \\ \hline
 $B \to J/\psi K_{S}$ &$40$\\ \hline
 $B \to \phi K_{S}$ &$0.6-2$\\ \hline 
 $ B \to \pi^{0} K_{S}$ & $0.02 - 0.4$ \\ \hline 
 $ B \to D^{0}_{CP} \pi^{0}$& $16$\\ \hline 
 $B \to D^{+} D^{-}$ & $30-50$\\ \hline 
 $B \to J/\psi \pi^{0}$ & $2$\\ \hline 
 $B \to \phi \pi^{0}$ &$1-4 \times 10^{-4}$ \\ \hline 
 $B \to K^{0} \bar{K}^{0}$ &$0.007-0.3$\\ \hline 
 $B \to \pi^{+} \pi^{-}$ &$0.2-2$\\ \hline 
 $B \to \pi^{0} \pi^{0}$ & $0.003-0.09$\\ \hline 
 $B \to K^{+} K^{-}$ & $< 0.5 $\\ \hline 
 $B \to D^{0} \bar D^{0}$ & $<20$ \\ \hline
 \end{tabular}
 \end{center}
 \end{table}
 
Coming to the SUSY contributions, we make use of the Wilson coefficients for
the gluino contribution (see eq. (12) of ref. \cite{GGMS}) and parameterize the
matrix elements as we did before for the SM case. We obtain the ratios of the
SUSY to the SM amplitudes as reported in table~\ref{tab:ampli} for $\tilde q$
and $\tilde g$ masses of 250 GeV and 500 GeV (second and third  row, 
respectively). From the table, 
one concludes that the inclusion of the various terms
in the amplitudes, $DE$, $DA$, etc., 
can  modify the ratio $r$ of  SUSY to SM contributions up to one
order of magnitude.

In terms of the decay amplitude $A$, the CP asymmetry reads 
\begin{equation}
{\cal A}(t) = \frac{(1-\vert \lambda\vert^2) \cos (\Delta M_d t )
-2 {\rm Im} \lambda \sin (\Delta M_d t )}{1+\vert \lambda\vert^2} 
\label{eq:asy}
\end{equation}
with $\lambda=e^{-2i\phi^M}\bar{A}/A$. 
In order to be able to discuss the results  model-independently,
we have labelled as $\phi^M$ the generic  mixing phase.
The ideal case occurs when  one  decay 
amplitude only appears in (or dominates)
a decay process: the CP violating asymmetry is  then determined by the
 total phase   $\phi^T=\phi^M+\phi^D$, where $\phi^D$  is the weak phase
  of the decay.
This ideal situation is spoiled by the presence of
several interfering amplitudes.
If the ratios $r$ in table~\ref{tab:ampli} are small, then the uncertainty on 
the sine of the CP phase is $<  r $, while if $r$ is O(1)  $\phi^T$
receives, in general,  large corrections.

The results of our analysis are summarized in tables~\ref{tab:BR} and
\ref{tab:results} which 
collect the branching ratios and CP phases  for the relevant $B$ decays of
table~\ref{tab:ampli}. $\Phi^D_{SM}$ denotes the decay phase in the SM; for each
channel, when two amplitudes with different weak phases are present, we
indicate the SM phase of the Penguin (P) and Tree-level (T) decay
amplitudes. The range of variation of $r$ in the SM ($r_{SM}$) is deduced from
table~\ref{tab:ampli}. For  $B \to K_S \pi^{0}$ the
 penguin contributions (with a vanishing phase) dominate over the
tree-level amplitude   because the latter is Cabibbo suppressed. 
For the channel $b
\to s \bar s d$  only penguin operators or penguin contractions of
current-current operators  contribute. The phase $\gamma$ is present in the
penguin contractions of the $(\bar b u)(\bar u d)$ operator, 
denoted as $u$-P $\gamma$
in table~\ref{tab:results}.\cite{Fleischer}  
 $\bar b d \to \bar q q $ indicates processes occurring via annihilation 
 diagrams which can be measured
 from the last two channels of table~\ref{tab:results}.
In the case $B \to K^{+} K^{-}$ both
current-current and penguin operators contribute. In $B \to D^{0} \bar
D^{0}$ the contributions
from  the $(\bar b u) (\bar u d)$ and the   $(\bar b c) (\bar c d)$
current-current operators   (proportional to the phase $\gamma$) 
tend to cancel out.

\begin{table}
 \begin{center}
 \caption[]{CP phases for $B$ decays. $\phi^{D}_{SM}$
 denotes the decay phase in 
 the SM; T and P denote Tree and Penguin, respectively; for each
 channel, when two amplitudes with different weak phases are present, 
 one is given in the first row, the other in the last one
 and the ratio of the two  in the $r_{SM}$ column. $\phi^{D}_{SUSY}$
 denotes the phase of the SUSY amplitude, and the ratio of the SUSY to SM
 contributions is given in the $r_{250}$ and $r_{500}$ columns for the
 corresponding SUSY masses.}
 \label{tab:results}
 \begin{tabular}{|ccccccc|}
 \hline 
 \ss{Incl. }&\ss{ Excl. }&\ss{ $\phi^{D}_{\rm SM}$ }&\ss{ $r_{\rm SM}$ }&\ss{ 
 $\phi^{D}_{\rm SUSY}$ }&\ss{ $r_{250}$ }&\ss{ $r_{500}$ }\\ 
 \hline
\ss{ $b \to c \bar c s$ }&\ss{ $B \to J/\psi K_{S}$ }&\ss{ 0 }&\ss{ -- }&\ss{
 $\phi_{23}$ }&\ss{ $0.03-0.1$ 
 }&\ss{$0.008-0.04$ }\\ \hline
 \ss{ $b \to s \bar s s$ }&\ss{ $B \to \phi K_{S}$ }&\ss{ 0 }&\ss{ -- }&\ss{
 $\phi_{23}$ }&\ss{ $0.4-0.7$ }&\ss{ 
 $0.09-0.2$ }\\ \hline \ss{
 $b \to u \bar u s$ }&\ss{ }&\ss{ P $0$ }&\ss{  }&\ss{  }&\ss{  }&\ss{
  }\\ 
\ss{}&\ss{$ B \to \pi^{0} K_{S}$} &\ss{  }&\ss{ $0.01-0.08$ }&\ss{
  $\phi_{23}$ }&\ss{ $0.4-0.7$ }&\ss{ 
 $0.09-0.2$ }\\ \ss{ 
 $b \to d \bar d s$ }&\ss{ }&\ss{ T $\gamma$ }&\ss{  }&\ss{  }&\ss{  }&\ss{
  }\\ \hline \ss{
 $b \to c \bar u d$ }&\ss{ }&\ss{ 0 }&\ss{  }&\ss{  }&\ss{  }&\ss{
  }\\ \ss{ 
  }&\ss{$ B \to D^{0}_{CP} \pi^{0}$ }&\ss{  }&\ss{ 0.02 }&\ss{ -- }&
  \ss{ -- }&\ss{
 -- }\\ \ss{  
 $b \to u \bar c d$ }&\ss{ }&\ss{ $\gamma$ }&\ss{  }&\ss{  }&\ss{  }&\ss{
  }\\ \hline \ss{
  }&\ss{ $B \to D^{+} D^{-}$ }&\ss{ T $0$ }&\ss{ $0.03-0.3$ }&\ss{  }&\ss{
  $0.007-0.02$ }&\ss{ 
  $0.002-0.006$ }\\ \ss{ 
  $b \to c \bar c d$}&\ss{ }&\ss{ }&\ss{  }&\ss{ $\phi_{13}$ }&\ss{ }&\ss{
  }\\ \ss{ 
  }&\ss{ $B \to J/\psi \pi^{0}$ }&\ss{ P $\beta$ }&\ss{ $0.04-0.3$ }&\ss{
  }&\ss{ $0.007-0.03$ }&\ss{ 
  $0.002-0.008$ 
  }\\ \hline \ss{
  }&\ss{ $B \to \phi \pi^{0}$ }&\ss{ P $\beta$}&\ss{ -- }&\ss{ }&\ss{
  $0.06-0.1$ }&\ss{ 
  $0.01-0.03$ }\\ \ss{ 
  $b \to s \bar s d$}&\ss{ }&\ss{  } &&\ss{ $\phi_{13}$ }&\ss{ }&\ss{
  }\\ \ss{ 
  }&\ss{ $B \to K^{0} \bar{K}^{0}$ }&\ss{ {\it u}-P
  $\gamma$  
  }&\ss{ $0-0.07$ }&\ss{ }&\ss{ $0.08-0.2$ }&\ss{
  $0.02-0.06$ 
  }\\ \hline \ss{
 $b \to u \bar u d$ }&\ss{ $B \to \pi^{+} \pi^{-}$ }&\ss{ T
 $\gamma$  
 }&\ss{ $0.09-0.9$ }&\ss{ $\phi_{13}$ }&\ss{ $0.02-0.8$ }&\ss{
 $0.005-0.2$ }\\ \ss{ 
 $b \to d \bar d d$ }&\ss{ $B \to \pi^{0} \pi^{0}$ }&\ss{
 P $\beta$ }&\ss{ $0.6-6$  
 }&\ss{ $\phi_{13}$ }&\ss{ $0.06-0.4$ }&\ss{
 $0.02-0.1$ }\\ \hline \ss{
 }&\ss{ $B \to K^{+} K^{-}$ }&\ss{ T $\gamma$ }&\ss{ $0.2-0.4$ }&\ss{ }&\ss{
 $0.04-0.1$ }&\ss{$0.01-0.03$ 
  }\\ \ss{
 $b \bar d \to q \bar q$ }&\ss{ }&\ss{ }&\ss{ }&\ss{ $\phi_{13}$}&
 \ss{ }&\ss{ }\\ \ss{
 }&\ss{ $B \to D^{0} \bar D^{0}$ }&\ss{ P $\beta$ 
 }&\ss{ only $\beta$ }&\ss{  }&\ss{ $0.01-0.03$ }&\ss{$0.003-0.006$ }\\  \hline
 \end{tabular}
 \end{center}
 \end{table}

SUSY contributes to the decay amplitudes with  phases 
induced by  $\delta_{13}$ and
$\delta_{23}$ which we denote as $\phi_{13}$ and $\phi_{23}$. The ratios of
$A_{SUSY}/A_{SM}$ for SUSY masses of 250 and 500 GeV as obtained from 
table~\ref{tab:ampli} are reported in the $r_{250}$ and $r_{500}$ columns 
of table~\ref{tab:results}. 

We now draw some conclusions from the results of table~\ref{tab:results}. 
In the SM, the first
six  decays  measure directly the mixing phase $\beta$, up to
corrections which, in most of the cases, are expected to be small. 
These corrections, due to the presence of  two 
amplitudes contributing with different phases,  produce 
 uncertainties of $\sim 10$\% in   $B \to K_S \pi^{0}$,
 and  of $\sim 30$\%  in $B \to D^{+} D^{-}$ and $B \to
J/\psi \pi^{0}$.   In spite
of the uncertainties,  however, there are cases where
 the SUSY contribution gives rise to significant changes. 
 For example, for SUSY masses of O(250) GeV, SUSY corrections  can  shift the
measured value of the sine of the phase in
 $B \to \phi K_S$ and in $B \to K_S \pi^{0}$ decays by an amount of 
 about 70\%.  For these decays  SUSY effects are sizeable even for 
masses of 500 GeV.  In $B \to
J/\psi K_S$  and $B \to \phi \pi^0$ decays, SUSY effects are only  about $10$\%
but SM uncertainties are negligible.  In $B \to K^0 \bar{K}^0$ 
the larger  effect, $\sim 20$\%,   is partially covered by the 
indetermination of 
about $10$\%  already existing in the SM. 
Moreover the rate for this channel is expected to be rather small.
In $B \to D^{+} D^{-}$  and $B \to K^{+} K^{-}$, SUSY effects are 
completely obscured  by the errors in the estimates of the SM amplitudes.
In $B^0\to D^0_{CP}\pi^0$ the asymmetry  is sensitive to the mixing angle 
$\phi_M$ only because the decay amplitude is unaffected by SUSY. 
This result can be used in connection with $B^0 \to K_s \pi^0$, since
a difference in the measure of the phase  is  a manifestation
of SUSY effects.
\par 
Turning to $B \to \pi \pi$ decays, both the uncertainties
in the SM  and  the SUSY contributions are very large. Here we
witness the presence of three independent amplitudes with different phases 
and of comparable size.  The observation of SUSY effects in
the $\pi^{0} \pi^{0}$ case is hopeless. The possibility of 
separating SM and SUSY contributions  by using the isospin
analysis remains an open possibility which deserves further investigation.
For a thorough discussion of the SM uncertainties in $B \to \pi \pi $ see
ref.~\cite{cfms} 

In conclusion, our analysis shows that measurements of CP asymmetries in
several channels may allow the extraction of the CP mixing phase and
to disentangle  SM and SUSY contributions to the CP decay phase.  
The golden-plated decays in this respect are $B \to \phi K_S$
and $B \to K_S \pi^0$ channels. The size of the SUSY effects is
clearly controlled by the the non-diagonal SUSY mass
insertions $\delta_{ij}$, which for illustration we have assumed to have the
maximal value compatible with the present experimental limits on 
$B^0_d$--$\bar B^0_d$ mixing.

\end{document}